\documentclass[a4paper,fleqn,usenatbib]{mnras}

\usepackage{graphicx}	
\usepackage{amsmath}	
\usepackage{amssymb}	
\usepackage{multicol}        
\usepackage{bm}		
\usepackage{pdflscape}	
\usepackage{color}

\title[]{Modified gravity (MOG) and the cluster Abell 1689 acceleration data}
\author[J .W. Moffat and M.H. Zhoolideh Haghighi]{J .W. Moffat$^{1 }$
	\thanks{zhoolideh_m@sharif.edu}, \ M.H. Zhoolideh Haghighi$^{1, 2}$\thanks{rahvar@sharif.edu}  \\
	$^1$ Perimeter Institute for Theoretical Physics, 31 Caroline St. N., Waterloo, ON, N2L 2Y5,Canada  \\
	$^{2 }$Department of Physics, Sharif University of Technology, P.O.
	Box 11155-9161, Tehran, Iran}

\pagerange{\pageref{firstpage}--\pageref{lastpage}} \pubyear{}
\def\LaTeX{L\kern-.36em\raise.3ex\hbox{a}\kern-.15em
	T\kern-.1667em\lower.7ex\hbox{E}\kern-.125emX}

\begin{document}
	\label{firstpage}
	\pagerange{\pageref{firstpage}--\pageref{lastpage}}

\maketitle

\begin{abstract}
The galaxy cluster system Abell 1689 has been well studied and yields good lensing and X-ray gas data. Modified gravity (MOG) is applied to the cluster Abell 1689 and the acceleration data is well fitted without assuming dark matter.  Newtonian dynamics and Modified Newtonian dynamics  (MOND) are shown not to fit the acceleration data, while a dark matter model based on the Navarro-Frenk-White (NFW) mass profile is shown to fit the acceleration data for the radial range $r > 200$ kpc.
\end{abstract}

\section{Introduction}

Zwicky in 1933 (\cite{Zwicky1,Zwicky2}) found that unseen mass had to be assumed to exist to explain the observational data for the Coma cluster. Subsequently, it was discovered that this unseen mass called dark matter had to be incorporated into gravity to explain the rotation curves of galaxies~(\cite{rubin2}). Underground experiments~(\cite{LUX,pandaX}) as well as astrophysical searches have failed to detect dark matter particles.  An alternative explanation for the missing dark matter is to modify Newtonian and Einstein gravity. One such fully covariant and Lorentz invariant alternative theory is Scalar-Tensor-Vector-Gravity (STVG), also known as MOG~(\cite{Moffat1}). MOG has passed successful tests in explaining rotation velocity data of spiral and dwarf galaxies~(\cite{MoffatRahvar1}),~(\cite{HaghighiRahvar}), globular clusters~(\cite{MoffatToth}) and clusters of galaxies~(\cite{MoffatRahvar2}). 

Recently, it was claimed~\citep{Nieuwenhuizen2} that no modified gravity theory can fit the Abell 1689 acceleration data without including dark matter or heavy (sterile) neutrinos. The cluster A1689 is important, for good lensing and gas data are available and we have data from 3 kpc to 3 Mpc.  We will show that MOND~\citep{milgrom} does not fit the A1689 acceleration data, while MOG does fit the acceleration data as does a dark matter model based on an NFW mass profile for the cluster radial range $r \gtrsim  200$ kpc. 

\subsection{The MOG Acceleration}

We will use the weak field approximation of the MOG theory at astrophysical scales to describe the dynamics of 
galaxies and clusters of galaxies. The point particle acceleration is given by~(\cite{Moffat1,Moffat2}):
\begin{eqnarray}
\label{MOGpointAcceleration}
a(r)_{\rm MOG}=- \frac{G_NM}{r^2}\Big[1+\alpha -\alpha \exp(-\mu r)(1+{\mu}r)\Big].
\label{pointacceleration}
\end{eqnarray}
Here, $G_N$ denotes Newton's gravitational constant and $M$ denotes the mass.  The parameter $\alpha$ determines the strength of attractive gravitation through $G=G_N(1+\alpha)$, while $\mu$ is the mass of the vector field $\phi_\mu$ that produces a repulsive gravitational force through the gravitational charge:
\begin{equation}
Q_g=\int d^3x J^0=\sqrt{\alpha G_N}M,
\end{equation}
where $J^\mu=(J^0,J^i)\,(i=1,2,3)$ is the current density matter source of the gravitational vector field $\phi_\mu$, $G_N$ is Newton's gravitational constant and $M$ is the source mass. The MOG acceleration (\ref{MOGpointAcceleration}) satisfies the (weak) equivalence principle.

For a distribution of matter, the acceleration is given by
\begin{equation}
\label{mogacceleration}
\begin{split}
\mathbf{a}(\mathbf x)= - G_N\int d^3x'\frac{\rho(\mathbf x')(\mathbf{x}-\mathbf{x'})}
{|{\mathbf x}-{\mathbf x'}|^3}\\
\times\Big[1+\alpha -\alpha \exp(-\mu|{\mathbf x}-{\mathbf x'}|)(1+{\mu}|{\mathbf x}-{\mathbf x'}|)\Big].
\end{split}
\end{equation}
The application of the theory to galaxies as well as clusters of galaxies determines the best fit values:
$\alpha = 8.89\pm 0.34$ and $\mu = 0.042\pm 0.004~{\rm kpc}^{-1}$~\citep{MoffatRahvar1,MoffatRahvar2}. 
 
The potential for a distribution of matter with density $\rho$ is given by:
\begin{equation}
\label{mogphi}
\begin{split}
\Phi_{\rm eff}(\mathbf x) = - G_N \int d^3x'\frac{\rho(\mathbf x')}{|\mathbf x-\mathbf x'|}
\big[1+\alpha-\alpha\exp(-\mu|\mathbf x-\mathbf x'|)\big].
\end{split}
\end{equation}

\section{A1689 Data}

In order to see if MOG fits the cluster A1689 data, we employ the normalized acceleration data given by Nieuwenhuizen in his Fig. 2~(\cite{Nieuwenhuizen2}).
Nieuwenhuizen has obtained the acceleration data by using A1689 strong and weak lensing data. In particular, the two-dimensional mass density $\Sigma(r)$ has been determined by Limousin et al.~(\cite{Limousin}). From X-ray observations the electron and gas mass density can be obtained. We use 59 data points with 12 data points from~(\cite{Limousin}) at radii between 3 kpc and 271 kpc, 20 data points from ~(\cite{2010ApJ...723.1678C}), which overlap with the data of Limousin et al. in the window between 40 kpc and 150 kpc, and 27 data points from~(\cite{2015ApJ...806..207U}), which cover the range between 125 kpc and 3 Mpc. The gas data is from Chandra X-ray observations with a total exposure time of 150 ks~(\cite{2010ApJ...713..491M}) and in our calculations we use a cored Sersic electron 
density profile:
\begin{equation}
n_{e}(r)=n^{0}_{e} \exp\Big[k_{g}-k_{g}\biggl(1+\frac{r^2}{R_{g}^2}\biggr)^{\frac{1}{2n_{g}}} \Big].
\end{equation}
From~\citep{Nieuwenhuizen2}, we obtain $n^{0}_{e}=0.0673\pm0.0027\,{\rm cm}^{-3}$,  $k_{g}=1.90\pm 0.20$,
$R_{g}=21.2\pm2.4\,{\rm kpc}$ and $n_{g}=2.91\pm 0.11$. For a typical Z=0.3 solar metallicity the gas density is~(\cite{2009EL.....8659001N}):
\begin{equation}
\rho_{g}(r)=1.167 m_{N}n^{3}_{e}(r),
\end{equation} 
where $m_N$ denotes the nucleon mass.

Background galaxies are far away from the cluster, so strong lensing can be treated as occurring due to mass projected onto the plane through the cluster center. The strong lensing yields data for the mass density along the line-of-sight:
\begin{equation}
\Sigma(r)=\int^\infty_{-\infty}dz\rho(\sqrt{r^2+z^2}).
\end{equation}

\section{MOG, MOND and NFW profile acceleration results}

We investigate the possibility of fitting the A1689 data with MOG, MOND and with a dark matter model using a NFW profile. Although there are many galaxies in a galaxy cluster, the galaxy mass density of A1689 is dominated by the brightest cluster galaxy (BCG) in the center of the cluster. We will use the BCG density profile proposed in~(\cite{2005MNRAS.356..309L}):
\begin{equation}
\rho_{\rm Galaxy}(r)= \frac{M_{cg}(R_{co}+R_{cg})}{2 \pi^2(r^2+R_{co}^2)(r^2+R_{cg}^2)},
\end{equation}
where $M_{cg}$ and $R_{cg} $ denote the mass and the radial size of the central galaxy, respectively, while $R_{co}$ denotes its core size. In our calculations, we use in our BCG mass density profile $R_{co}=5$ kpc, $R_{cg} =150$ kpc and $M_{cg}\sim  10^{13} M_{\odot}$ from~(\cite{Nieuwenhuizen2}).

Although there is some observational evidence for A1689 possessing triaxial symmetry~\citep{2015ApJ...806..207U},  we assume similar to Nieuwenhuizen that A1689 has spherical symmetry in order to integrate the density profile and calculate the gravitational acceleration. It can be demonstrated that the assumption of spherical symmetry will not affect our results significantly~ \citep{Nieuwenhuizen2}. We will use the MOG point mass radial acceleration equation (\ref{MOGpointAcceleration}), where $M = M_{\rm bar}$ and $M_{\rm bar}$ denotes the baryon mass. By using 
\begin{equation}
\rho(r)=\rho_{gas}(r)+\rho_{\rm Galaxy}(r),
\end{equation}
and integrating $\rho(r)$ over the volume of the cluster, we obtain the mass $M_{\rm bar}(r)$. Because there is a dynamical differential equation for $\mu$, we can infer that $\mu$ is not a universal constant and it can vary for different physical systems~\citep{Moffat1,GW}:
\begin{equation}
\Box \mu= \frac{1}{G}\partial^{\alpha}G\partial_{\alpha}\mu+\frac{2}{\mu}\partial^{\alpha}\mu\partial_{\alpha}\mu+ \mu^2G \frac{\partial V(\phi_\mu)}{\partial \mu},
\end{equation}
where $\Box=\nabla^\alpha\nabla_\alpha$ and $V(\phi_\mu)$ is a potential that depends on both the vector field $\phi_\mu$ and $\mu$. We will choose different $\mu$ parameter values to fit the acceleration data. One choice is: $\alpha=8.89$ and $\mu=0.042\,{\rm kpc}^{-1}$ or $\mu^{-1}=24$ kpc, while the other choice is: $\alpha=8.89$ and $\mu=0.125\,{\rm kpc}^{-1}$ or $\mu^{-1}=0.8$ kpc. The first choice coincides with the fits to galaxy rotation curves~\citep{MoffatRahvar1}, while the second choice is the best fit for MOG and the observed A1689 acceleration data.

We have plotted the MOG predicted acceleration in Fig.1. The acceleration data is from~(\cite{Nieuwenhuizen2}). The acceleration curve for Newtonian gravity without dark matter, determined by $a_{\rm Newt}=G_NM_{\rm bar}/r^2$, is depicted in Fig.1. and does not fit the normalized acceleration data. In the case of MOND, we use the following interpolating function $\nu(a/a_0)$ to calculate the acceleration:
\begin{equation}
\nu\biggl(\frac{a}{a_{0}}\biggr)=\biggl[1+\biggl(\frac{a_{0}}{a}\biggr)^2\biggr]^{-\frac{1}{2}}.
\end{equation}
This leads to the MOND acceleration:
\begin{equation}
a_{\rm MOND}=\frac{a_{\rm Newt}}{\sqrt{2}} 
\biggl[1+\biggl(1+\biggl(\frac{2a_0}{a_{\rm Newt}}\biggr)^2\biggr)^{1/2}\biggr]^{1/2}.
\end{equation}
The predicted MOND normalized acceleration curve shown in Fig.1, follows closely the Newtonian acceleration curve without dark matter.
 
To obtain the A1689 dark matter normalized acceleration, we adopt the NFW profile~(\cite{NFW}):
\begin{equation}
\rho(r)_{\rm NFW}=\frac{\rho_{0}}{\frac{r}{r_{s}}\biggl(r+\frac{r}{r_{s}}\biggr)^2}.
\end{equation}
Let us consider the following relationships between concentration, $ R_{200},r_{s} $ and $M_{200}$:
\begin{equation}
C=\frac{R_{200}}{r_{s}},
\end{equation}
\begin{equation}
R_{200}=\biggl(\frac{M_{200}}{4/3\pi 200\rho_{crit}}\biggr)^{1/3},
\end{equation}
\begin{equation}
\rho_{0}=\rho_{crit}\delta_{0},
\end{equation}
\begin{equation}
\delta_{0}=\frac{200}{3}\frac{C^3}{\ln(1+C)-C/(1+C)}.
\end{equation}
Here, $\rho_{crit}$ is the critical density ($\rho_{crit}=277.3h^2 M_{\odot}{\rm kpc}^{-3}$ ) and $M_{200}$ is assumed to be the total mass of the dark matter halo.
The dark matter model acceleration is given by
\begin{equation}
a(r)_{\rm NFW}=- G_N\frac{M_{\rm dm}(r)}{r^2},
\end{equation}
where $M_{\rm dm}$ denotes the dark matter mass. 

The cluster A1689 is well-studied and different measured NFW profile parameters have been obtained, so we use the best fit parameters of three different groups for the sake of generality. We have from~\citep{2008MNRAS.386.1092L}: $\rho_{0}=9.6\pm 1.8\times 10^{-25}h^2 {\rm gr/cm}^3$ and $r_{s}=175\pm 18\,h^{-1} {\rm kpc}$, from \citep{broadhurst} $r_{s}=310^{+140}_{-120} {\rm kpc}/h$ and $C_{200}=6.5^{1.9}_{-1.6}$ and from \citep{2015ApJ...806..207U} $C_{200}=8.9\pm1.1$ and $M_{200}=(1.3\pm0.11)\times10^{15}M_\odot h^{-1} $ . After integrating the NFW profile and calculating $M_{\rm dm}(r)$, we obtain the acceleration curves depicted in Fig.1. We have normalized the acceleration in all cases to $a_0=cH_0/2\pi=1.14 \times 10^{-10}\, {\rm ms}^{-2}$.  These results are shown in Fig.1. 

In order to have a qualitative comparison of the goodness of fits of the mentioned models, we have calculated the $\chi^2$ for separate sets of data and we have reported the results in Table(\ref{Table1}). Because there are data points from different groups, we have calculated the $\chi^2$ separately for the different data points. The MOND prediction is the same as the Newtonian prediction, and MOND cannot fit the data (see Table (\ref{Table1})), except when dark matter such as heavy sterile neutrinos are included. By using the NFW parameters reported by \citep{2015ApJ...806..207U}, we can obtain a good fit to the Umetsu shear acceleration data ($\chi^2/N_{n.d.f}$= 1.23) for $r > 200$ kpc, but a poorer fit to the Umetsu WL data ($\chi^2/N_{n.d.f}$= 2.9) and the Coe SL data ($\chi^2/N_{n.d.f}$= 15.11). In the case of MOG, for the parameter choices $\alpha=8.89$ and $\mu=0.042\, {kpc}^{-1}$, we obtain a good fit for both the Umetsu shear data ($\chi^2/N_{n.d.f}$= 1.38) and the Umetsu WL data ($\chi^2/N_{n.d.f}$= 1.11) for $r > 50$ kpc, but not a good fit for the Coe SL data ($\chi^2/N_{n.d.f}$= 5.06). For the parameter choices $\alpha=8.89$ and $\mu=0.125\, {kpc}^{-1}$, we obtain a good fit for both the Umetsu shear data ($\chi^2/N_{n.d.f}$= 1.38) and the Umetsu WL data ($\chi^2/N_{n.d.f}$= 1.09) and not a bad fit for the Coe SL data ($\chi^2/N_{n.d.f}$= 2.012). None of the models that we have presented fit the data of Limousin et al. for small 
radius $r\lesssim 10$ kpc.

\section{Conclusions}
 
The lensing and the Chandra X-ray data can determine the total baryon density for the cluster A1689. The acceleration for Newtonian, MOND and MOG models is derived from the baryon mass $M_{\rm bar}(r)$ by integrating the total baryon density, assuming that the BCG is the dominant mass in the A1689 cluster. The normalized acceleration for these models is then compared with acceleration data for A1689. The Newtonian and MOND predictions without dark matter fall far short of fitting the data at all values of the radial distance $r$.  A dark matter model based on a NFW density profile with the parameters $\rho_0$ and $r_s$ fits the Umetsu shear and WL acceleration data with the goodness of the fit reported in Table (\ref{Table1}) for $r > 200$ kpc. The MOG predicted acceleration without dark matter provides a good fit for the Umetsu WL and Umetsu shear data and not a bad fit for  Coe SL data with the $\chi^2$ reported in Table (\ref{Table1}) for $r\gtrsim 10$ kpc.

Modified gravity theories have been formulated to avoid the need for dark matter. Experiments have failed to detect dark matter particles motivating the study of alternative gravitational theories. The fully covariant and Lorentz invariant MOG theory fits galaxy dynamics data and cluster data. It also fits the merging clusters Bullet Cluster and the Train Wreck Cluster (Abell 520) without dark matter~(\cite{BrownsteinMoffat,IsraelMoffat}).  A MOG application to cosmology without dark matter can explain structure growth and the CMB data~(\cite{Moffat1,TothMoffat}). The fitting of the cluster A1689 data adds an important success for MOG as an alternative gravity theory without dark matter.
 
\section*{Acknowledgments}

We thank Martin Green, Niayesh Afshordi and Theodorus Nieuwenhuizen for helpful discussions. We thank Marceau Limousin for supplying data and for helpful discussions. This research was supported in part by Perimeter Institute for Theoretical Physics. Research at Perimeter Institute is supported by the Government of Canada through the Department of Innovation, Science and Economic Development Canada and by the Province of Ontario through the Ministry of Research, Innovation and Science.

\onecolumn

\begin{table}
\caption{Normalized $\chi^2 $ for the acceleration of different theories for different data sets. }
\begin{tabular}{ |c||c|c|c|c|c|c|  }
 \hline
 \multicolumn{6}{|c|}{$\chi^2/N_{n.d.f}$} \\
 \hline
  Data source & MOND  & MOG($\mu =$)  & MOG($\mu =$)  & NFW with parameters  & NFW with parameters & NFW with parameters\\
&  & $ 0.042_{-0.004}^{+0.004} $ & $0.125_{-0.06}^{+0.37}$   &from Lemze et al. & from Broadhurst et al. & from Umetsu et al. \\

 \hline
Coe et al.(2010)  SL  & 50.30    & 5.06 & 2.012  &   29.65 & 20.24 & 15.11\\
Umetsu et al. (2015) WL &6.99 & 1.11 & 1.09 &  2.28 & 4.44 & 2.90\\
Umetsu et al. (2015) $g_{t}$   &18.8 &  1.38 & 1.38 &   4.7 & 3.45 & 1.23\\
Limousin et al. (2007) $g_{t}$   &67.31 & 18.89 &  8.48 & 49.21 & 38.51 & 28.83\\
\hline
\end{tabular}
\label{Table1}
\end{table}
\begin{figure}
\centering \includegraphics{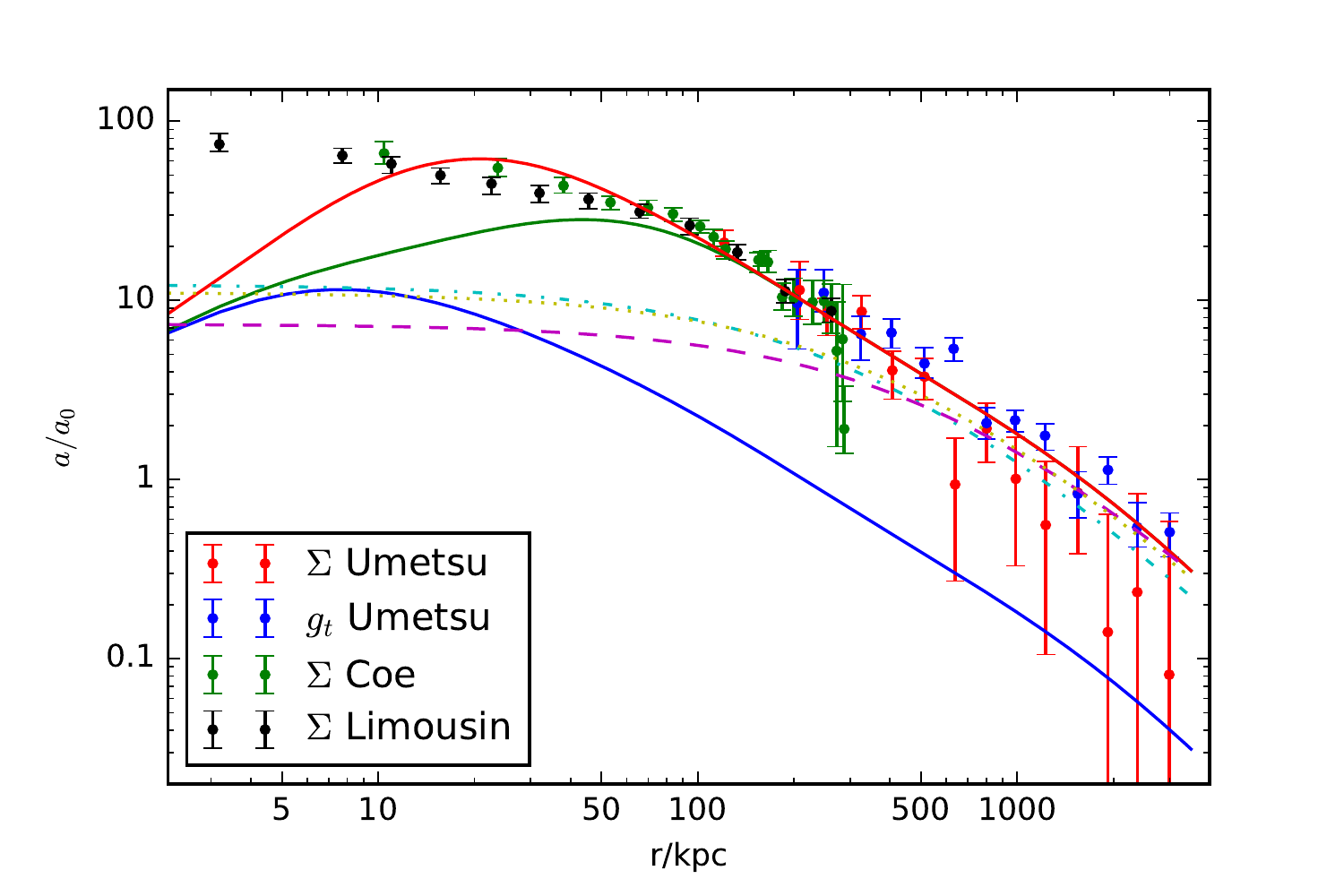}
\caption{The full red curve is the MOG prediction for acceleration with $\alpha=8.89$ and $\mu=0.125\,{\rm kpc}^{-1}$,The full green curve is the MOG prediction for acceleration with $\alpha=8.89$ and $\mu=0.042\,{\rm kpc}^{-1}$, the dash-dotted  curve is the NFW prediction with $\rho_{0}=9.6\times 10^{-25}h^2 {\rm gr}/{\rm cm}^3$  and $r_{s}=175\pm 18\,h^{-1} {\rm kpc}$ from~\citep{2008MNRAS.386.1092L}, the dashed  curve is for NFW  from \citep{broadhurst}  with $r_{s}=310^{+140}_{-120} kpc/h$ and $C_{200}=6.5^{1.9}_{-1.6}$, the dotted  curve is for NFW from \citep{2015ApJ...806..207U} with $C_{200}=8.9\pm1.1$ and $M_{200}=(1.3\pm0.11)\times10^{15}M_\odot h^{-1} $ and the full blue curve is the MOND and Newtonian predictions.}{\label{Fig.1.A1689}}
\end{figure}
\twocolumn

\end{document}